# Photodynamically Active Electrospun Fibres for Antibiotic-Free Infection Control

Amy Contreras,[1] Michael J. Raxworthy,[1,2] Simon Wood,[3] Jessica D. Schiffman,[4] Giuseppe Tronci[3,5] *

[1] Institute of Medical and Biological Engineering, University of Leeds, Leeds, LS2 9JT, UK

[2] Neotherix Ltd., The Hiscox Building, Peasholme Green, York, YO1 7PR, UK

[3] School of Dentistry, University of Leeds, Leeds, LS2 9JT, UK

[4] Department of Chemical Engineering, University of Massachusetts Amherst, 240 Thatcher Rd, Amherst MA 01003-9364, USA

[5] Clothworkers' Centre for Textile Materials Innovation for Healthcare, School of Design, University of Leeds, Leeds, LS2 9JT, UK

* Email correspondence: g.tronci@leeds.ac.uk (G.T.)

**Graphical Abstract**

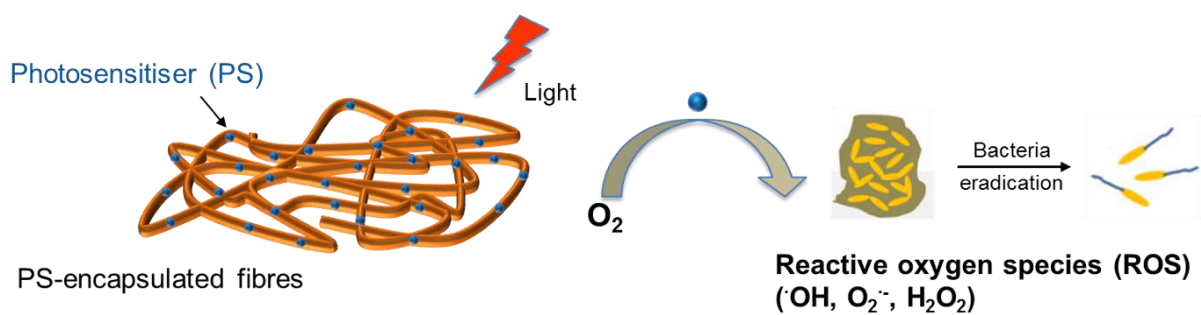

**Abstract**

Antimicrobial biomaterials are critical to aid in the regeneration of oral soft tissue and prevent or treat localised bacterial infections. With the rising trend in antibiotic resistance, there is a pressing clinical need for new antimicrobial chemistries and biomaterial design approaches enabling on-demand activation of antibiotic-free antimicrobial functionality following an infection that are environment-friendly, flexible and commercially-viable. This study explores the feasibility of integrating a bioresorbable electrospun polymer scaffold with localised antimicrobial photodynamic therapy (aPDT) capability. To enable aPDT, we encapsulated a photosensitiser (PS) in polyester fibres in the PS inert state, so that the antibacterial function would be activated on-demand via a visible light source. Fibrous scaffolds were successfully electrospun from FDA-approved polyesters, either poly(*ε*-caprolactone (PCL) or poly[(*rac*-lactide)-co-glycolide] (PLGA) with encapsulated PS (either methylene blue (MB) or erythrosin B (ER)). These were prepared and characterised with regards to their loading efficiency (UV-Vis spectroscopy), microarchitecture (SEM, porometry and BET (Brunauer-Emmett-Teller) analysis), tensile properties, hydrolytic behaviour (contact angle, dye release capability, degradability) and aPDT effect. The electrospun fibres achieved an ~100 wt.% loading efficiency of PS, which significantly increased their tensile modulus and reduced their average fibre diameter and pore size with respect to PS-free controls. *In vitro*, PS release varied between a burst release profile to limited release within 100 hours depending on the selected scaffold formulation, whilst PLGA scaffolds displayed significant macroscopic shrinkage and fibre merging following incubation in phosphate buffered saline solution. Exposure of PS-encapsulated PCL fibres to visible light successfully led to at least a 1 log reduction in *E. coli* viability after 60 minutes of light exposure whereas PS-free electrospun controls did not inactive microbes. This study successfully demonstrates the significant potential of PS-encapsulated electrospun fibres as photodynamically active biomaterial for antibiotic-free infection control.

## 1. Introduction

Regenerative medical devices are required in dental applications to aid in the regeneration of tissue in the oral cavity [1]. A commonly occurring issue with oral surgery is bacterial infection as the oral microflora is known to be complex and diverse [1,2]. This leads to graft loss and the need for additional surgical interventions [3]. The current treatment for these infections relies on oral antibiotics but, due to the alarming rise in antimicrobial resistance (AMR), other methods of treatment are being explored [4].

Antimicrobial photodynamic therapy (aPDT) is an alternative treatment, which uses a photosensitiser (PS) to kill bacterial cells locally following the application of convenient light sources without the need to administer antibiotics [5]. PDT originates from observations made over 100 years ago suggesting that the combination of light in the presence of molecular oxygen and harmless dyes can lead to the death of microorganisms. It has since been used throughout medical science, predominantly for the treatment of cancer but also in dermatology and eye disorders [5,6]. More recent applications have moved back towards the original use of PDT by using PS to target infections locally. The key advantages of aPDT is that it kills all bacteria, regardless of resistant strains, without further induction of resistance triggered [7]. Another benefit of aPDT is that both the PS and the light are non-toxic alone, but when combined in the correct dosimetry they can be tuned to be toxic to all organisms including viruses, virulence factors, fungi and bacteria within a localized region[8].

In a non-activated PS, a pair of electrons in a molecular orbital exist in the PS ground state. To activate the PS, light of a specific wavelength needs to be applied. This is normally light in the visible or near-infrared region of the electromagnetic spectrum and is PS-specific[9]. This light provides the energy to excite one of these ground state electrons into an activated state, where it can give rise to photochemical reactions, i.e. Type I PDT and Type II PDT. Each of these mechanisms occurs concurrently, but the type and the local environment of the PS *in vivo* affects the ratio of the two[8]. Type I PDT reactions are commonly associated with the killing of microbes and involve electron transfer reactions that generate hydroxide ions. These highly reactive species initiate radical chain reactions with fatty acids, cholesterols and lipids, resulting in cell death. Type II reactions are typically used to target cancer cells by producing singlet oxygen ($^1O_2$), a form of oxygen where the spins of the two unpaired electrons that occupy two degenerate molecular orbitals are opposed to each other. This electronically excited state is significantly more reactive than the lower energy electronic ground state of triplet oxygen ($^3O_2$), in which the spins of the unpaired electrons are aligned[10].

Although commonly applied for cancer therapy[11], the use of aPDT in dental surgery accounts for the largest growth of aPDT in clinical infection treatment [5,6] because PS such as methylene blue (MB) and erythrosin B (ER) have proved effective against bacterial biofilms and for treating oral infections[12]. The inexpensive and quick-to-use treatment method suits dental surgery well because the oral cavity is easily accessed using a light source by the dental practitioner, which allows specific targeting of only the affected areas of the mouth and the dental light is in the absorbance range of MB (wavelengths: 605 and 665 nm) and ER (wavelength: 529 nm)[13].

MB has recently been loaded on to keratin films to provide a localised and long-term inhibition of *S. aureus,* reaching a bacterial killing rate of 99.9% after 75 minutes light activation[14]. MB has also been shown to reduce the viability of microorganisms in biofilms grown on discs of bovine dentin upon LED light activation[15]. Similarly, ER-loaded nanoparticles exhibited significantly higher antimicrobial efficacy against *S. aureus* cells than pure PS in solution[16]. Given their porous microarchitecture, large surface area and biomimetic fibrous characteristics, we suggest that further improvements in the delivery of PS could be achieved using fibrous scaffolds, especially to integrate sustained antimicrobial activity with soft tissue repair capability. Polyesters, including poly(ε-caprolactone) (PCL) and poly[(*rac*-lactide)-co-glycolide] (PLGA), are commonly used in regenerative medicine applications due to their hydrolytic degradability, fine-controllability of macroscopic properties[17]. They are cost-effective and versatile as they can be highly tuned to the intended application[18]. Electrospinning is a facile technique used to efficiently produce micro- to nano-scale fibres of polymer solutions at a high production rate with low associated costs[19]. Although the principle of the technique was discovered over 120 years ago[20], only within the past 15 years has it become widespread due to the straightforward manufacture, enhanced scalability and the increased interest in nanoscience and tissue engineering[21]. Consequently, there are many papers which focus on the specifics of the electrospinning process[22,23]. The diameter of fibres produced by electrospinning are smaller than most other techniques such as melt spinning, wet spinning, self-assembly and phase separation[21,24]. The pore sizes within the fibrous structure can be useful for cell infiltration and release of incorporated drugs[21]. The three-dimensional porous structure integrated within the scaffold also mimics the extracellular matrix of biological tissues, such as the oral mucosa[25], making it ideal for oral tissue repair applications.

In this study, we investigated whether polyester-based fibrous scaffolds could be encapsulated with clinically-approved PS, i.e. MB and ER, to generate on-demand antimicrobial effect following aPDT principles. We hypothesised that electrospinning would offer a suitable one-step manufacturing route to generate scaffolds, avoiding time-consuming solution-to-fibre and fibre-to-fabric manufacturing steps. We selected two FDA-approved building blocks, i.e. PCL and PLGA, in light of their fibre-forming capability and to comply with current regulatory medical device framework[26]. The use of FDA-approved polymers in the scaffold prevents the expensive and timely process of applying for approval for a new polymer[27]. The effect of polymer building block as well as PS type and dosage on electrospinning solution, scaffold microarchitecture, tensile properties and antimicrobial effect was investigated, aiming to achieve aPDT-equipped fibres with reliable structure-property-function relationships.

## 2. Materials and Methods

### 2.1 Preparation of electrospinning solutions

1,1,1,3,3,3-hexafluoro-2-propanol (HFIP) solvent was sourced from Fluorochem Ltd. PCL ($M_n$: 80,000 g·mol$^{-1}$) was sourced from Sigma Aldrich, whilst PLGA ($M_n$: 63,000 g·mol$^{-1}$, 75:25 molar ratio of lactic and glycolic acid units) was purchased from Purac Biomaterials (PURASORB® PDLG 7507). PCL was used at 6% w/w and PLGA at 12% w/w concentration in HFIP. Methylene Blue (molecular mass of 319.85 g·mol$^{-1}$) and Erythrosin B (molecular mass of 835.90 g·mol$^{-1}$) dyes were both sourced from Sigma Aldrich. They were each used at a concentration of 2.2 mM in the electrospinning solution. For the bactericidal study, additional scaffolds with double the concentration of dye (4.4 mM) were produced. Polymer, dye and HFIP were weighed together into sealed flasks and covered with foil to protect from ambient light. They were stirred at room temperature for 48 hours in foil-covered containers to allow for dissolution of all components.

### 2.2 Measurements of solution viscosity and surface tension

Viscosity measurements were taken at room temperature using a Brookfield DV-E bench top viscometer (Brookfield Engineering Laboratories, Inc., Middleboro, MA, USA) at varying shear stresses with 9.0 ml of solution and spindle 31. Electrospinning solutions were also characterised as per their surface tension. Density of solutions were calculated by weighing 1 ml of solution in triplicate for each solution and plotting mass versus volume, with the densities calculated as 1.52±0.04 g/ml. Each electrospinning solution was loaded into a 2ml syringe with 18-gauge blunt-ended needle. The solution was ejected until a stable droplet was formed at the needle tip. KSV Pendant Drop equipment was used with Attension Theta Software Version 4.1.9.8 to analyse the droplet with 60 images taken over one minute.

### 2.3 Electrospinning

Polymer solutions were transferred into a 10 ml plastic syringe with an 18-gauge blunt-ended needle, which was then loaded into a syringe pump. A pump rate of 0.03 ml·min$^{-1}$ was used with an applied voltage of 16 kV. A cylindrical grounded mandrel (height = 125 mm, diameter = 75 mm) was coated in aluminium foil at 100 mm distance away from the needle tip and rotated at 30 rpm. Scaffolds were electrospun for 55±5 minutes in dark conditions. Scaffolds were dried under reduced pressure in a vacuum desiccator for 72 hours to remove residual solvent. Scaffolds sealed in foil packets and frozen until use to prevent degradation.

**2.4 Loading Efficiency (LE)**

Samples were cut into 1 cm diameter round discs and weighed individually on an analytical balance. These were then incubated in glass vials with 5 ml of HFIP and rolled at 60 rpm for 48 hours to dissolve. A standard UV-vis calibration curve was drawn with the PS dissolved in HFIP over an appropriate concentration range using a photometric plate reader. The *LE* value was calculated for each PS according to Equation 1:

$$LE = \frac{m_d}{m_e} \times 100 \qquad \textbf{(Equation 1)}$$

where $m_d$ and $m_e$ are the determined and expected values of PS mass loaded in the electrospun scaffold, respectively.

**2.5 Scanning electron microscopy (SEM)**

Dry samples were attached to metal stubs using carbon double-sided stickers and sputter coated with gold twice before being analysed on a 6-sample multi-stub holder on a Hitachi Scanning Electron Microscope at 4000x magnification. Scaffolds were sensitive to high vacuum settings so VP-SEM (variable pressure-SEM) low vacuum setting was used (270 Pa). Randomly selected locations were chosen to capture five images of each scaffold, and ten fibre diameters were taken from each image.

### 2.6 Brunauer-Emmett-Teller (BET) analysis

Micrometrics FlowPrep 060 was used to flush samples (approximately 0.4 g) with $N_2$ at 40 ºC for 4 hours prior to analysis. Micrometrics TriStar 3000 Surface Area and Porosity Analyzer used along with complementary Tristar 3000 software to analyse the samples with a full isotherm being produced for each sample.

### 2.7 Porometry

Samples were soaked in a low surface tension Galpore125 (perfluoroether, surface tension 15.6 mN·m$^{-1}$) solution before being displaced with air at a specific pressure within the POROLUX™ 100FM porometer. The Young-Laplace equation was used to convert this pressure into the diameter of the capillary (Equation 2). The associated POROLUX™ software was used to calculate the largest and smallest pores, mean flow pore size, and the distribution of pore sizes in the scaffold. The pore diameter was calculated as follows:

$$\text{Pore diameter} = \frac{4 \times \cos\theta \times \gamma}{P} \qquad \textbf{(Equation 2)}$$

with $P$ representing the pressure required to displace the liquid from pore, $\theta$ representing the contact angle of the wetting fluid with the scaffold and $\gamma$ representing the surface tension of Galpore125[28].

### 2.8 Water Uptake Analysis

Samples were cut into 1 cm$^2$ squares and weighed individually before being incubated in well plates at 37°C in 3 ml of distilled water ($_d$H$_2$O) for 24 hours. Samples were then removed and blotted dry on filter paper to remove non-bonded water before being weighed again on an analytical balance. The percentage water uptake of each scaffold was calculated according to Equation 3:

$$\text{Water uptake} = \frac{m_w - m_d}{m_d} \times 100 \qquad \textbf{(Equation 3)}$$

Where $m_w$ and $m_d$ represent the mass values of hydrated and dry scaffold discs. All samples were analysed in triplicate.

## 2.9 Contact Angle measurements

Static contact angle measurements were recorded in triplicate for each scaffold using a FTA 4000 Contact Angle Goniometer and the associated software package. The scaffolds were attached to glass slides to hold them flat for analysis. A microsyringe was used to drop deionised water onto the surface of the scaffold. After a few seconds, an image was taken of the droplet before analysis was performed on the shape of the droplet in image. For the film analysis, glass slides were coated in the corresponding electrospinning solution and left for the HFIP to evaporate for 7 days. Following this time, the films were analysed in the same way.

## 2.10 Hydrolytic degradation tests

Samples were cut into 1 cm$^2$ squares and weighed before being incubated with 5 ml of phosphate buffered saline (PBS) solution in sealed falcon tubes at 37°C for up to 8 weeks. At selected time points, samples were removed, washed in distilled water three times for 5 minutes each time on a shaker plate and blotted dry before being dried in vacuum desiccator for 1 week. All samples were analysed in triplicate. The percentage mass loss of the scaffolds was calculated according to Equation 4:

$$Percentage\ Mass\ Loss = \frac{m_d - m_t}{m_d} \times 100 \qquad \text{(Equation 4)}$$

where $m_t$ and $m_d$ represent the mass values of either the dry partially-degraded scaffold disc at the selected time point *t*, or the dry, freshly-prepared electrospun scaffold disc, respectively.

### 2.11 Release kinetics study

Samples were cut into discs and weighed (ca. 20 mg) before being incubated with 5 ml of PBS solution at 37°C for up to 5 hours. At selected time points, 100 µl of the solution was collected, analysed by UV-Vis spectroscopy, and added back to the sample. The collected solutions (100 µl) were analysed on a microplate reader to record peak absorbance at either 605 nm (for MB) or 529 nm (for ER). Resulting absorbance values were converted into concentration of PS in the medium via a linear absorbance-concentration calibration curve ($R^2 > 0.99$) obtained by measuring PS solutions prepared in PBS whose PS concentrations covered the range used for scaffold PS encapsulation.

### 2.12 Tensile tests

Dry scaffolds were cut into 10x30 mm strips and clamped into a James Heal™ Titan5 Universal Strength Testing machine with a 100 N load cell and T27 jaw scheme. The equipment was used with TestWise 2017 test analysis software. A pretension of 0.5 N was applied to the material, and then the material was elongated at a speed of 100 mm·min$^{-1}$ until the material failed. Force against elongation measurements were recorded for each sample five times. Stress-strain curves were plotted, and the elastic modulus calculated as the slope of the linear region of the curve. The toughness was measured as the integral under the stress-strain curve.

**2.13 Bacterial cell culture**

An overnight culture of bacteria was prepared. Briefly, 5 ml of Luria-Bertani (LB) media and 5 µl Carbenicillin (Carb) antibiotic were added to an autoclaved test tube in aseptic conditions. A pipette tip was flame sterilised and used to collect one colony from an agar plate containing the K12 MG1655 *E. coli* bacterial strain before being added to the test tube, which was again flame sterilised and sealed. This was incubated on a stirrer plate at 250 rotations per minute overnight at 37°C (approximately 16 hours).

When ready to use, a sample of the culture solution was diluted in a cuvette with additional LB media to obtain an absorbance reading at 600 nm between 0.5 and 1 against an LB media background. An absorbance reading of 1 was taken to be equivalent to $1.6 \times 10^8$ cells·ml$^{-1}$ according to the McFarland 0.5 standard to calculate the number of bacterial cells in the overnight culture solution[29]. A final bacterial cell concentration of approximately $2.5 \times 10^7$ cells·ml$^{-1}$ was used in each experiment.

**2.14 Antibacterial photodynamic therapy tests**

Room temperature scaffolds were cut into round discs with a diameter of 1.27 cm before being sterilised for 15 minutes on each side using an ultraviolet light source. These were then placed in a 6-well plate with 5ml of M9 minimal salts media per well. The well plates containing media and scaffolds were incubated at 37°C for 2 hours to allow for dye release. After this time, overnight bacterial culture solution and 5 µl of Carb was added. The plates were then irradiated (at 1 cm distance) with the light source (3500 lumen Husky LED portable work light) at 37°C for either 30, 60, or 120 minutes. During each experiment, a duplicate plate was wrapped in foil and placed in the same incubator as a 'dark' control to measure the level of toxicity of the PS and scaffold when not activated by the light source. The temperature of the incubator was monitored to ensure that there was no significant increase of temperature for the duration of the experiment. The *E. coli* K12 MG1655 bacteria used were engineered to fluoresce with a maximum excitation wavelength of 488nm and a maximum emission wavelength of 510nm thus eliminating the need to stain the bacteria with a 'live' stain. PI (propidium iodide), was used to monitor the number of dead cells in each experiment. Following light exposure, the bacterial solution was removed and 2 ml of PI solution (12.5 µl per ml of deionised water) was added to each well. This was left to incubate at room temperature for 15 minutes to allow bacterial staining to occur. After this, each sample was removed, rinsed in deionised water to remove excess stain and blotted gently on Kimwipes to remove excess water. The samples were then placed onto a glass slide and imaged directly under a Zeiss epifluorescence microscope using GFP (488 nm) and PI (535 nm) wavelength filters and ZenPro software. Images were taken at 20x magnification in three randomly chosen areas across the scaffold. ImageJ was used with the multi-point tool to count the live or dead cells on each image. The average log reduction in live bacteria was calculated according to Equation 5:

$$Log\ Reduction\ in\ Live\ Bacteria = \log_{10} \frac{n_{dead} + n_{live}}{n_{dead}} \qquad \textbf{(Equation 5)}$$

where $n_{dead}$ and $n_{live}$ are the number of dead (red) and live (green) bacteria, respectively, measured in the epifluorescence microscope image.

**2.15 Statistical Analysis**

Significant differences in the results were evaluated using an unpaired student's *t*-test. Data was deemed to be significantly different at *p* < 0.05. All data have been collected in triplicates and presented as Mean ± Standard Deviation.

**3. Results and Discussion**

The manufacturing and characterisation of a prototype regenerative medical device integrated with aPDT capability was successfully fabricated via electrospinning of a PS-loaded polyester solution (**Figure 1 A**). Our concept to achieve antibiotic-free, localised aPDT effect was that fibre-encapsulated PS, either MB or ER (**Figure 1 B-C**), will be released from the electrospun scaffold allowing PS uptake by bacterial cells. In line with clinical treatment decisions, light of a specific frequency can be introduced to activate the PS to generate toxic reactive species and selectively kill any bacteria present in tissue infected areas.

Sample nomenclature is as follows: samples of either fibres or electrospinning solutions were coded as XXX-YY, whereby XXX identifies the type of polymer, i.e. either PCL or PLGA **(Figure 1 D-E)**, whilst YY indicates the PS encapsulated in the sample, either MB or ER. Control samples without PS will be called either PCL-ND or PLGA-ND throughout the results.

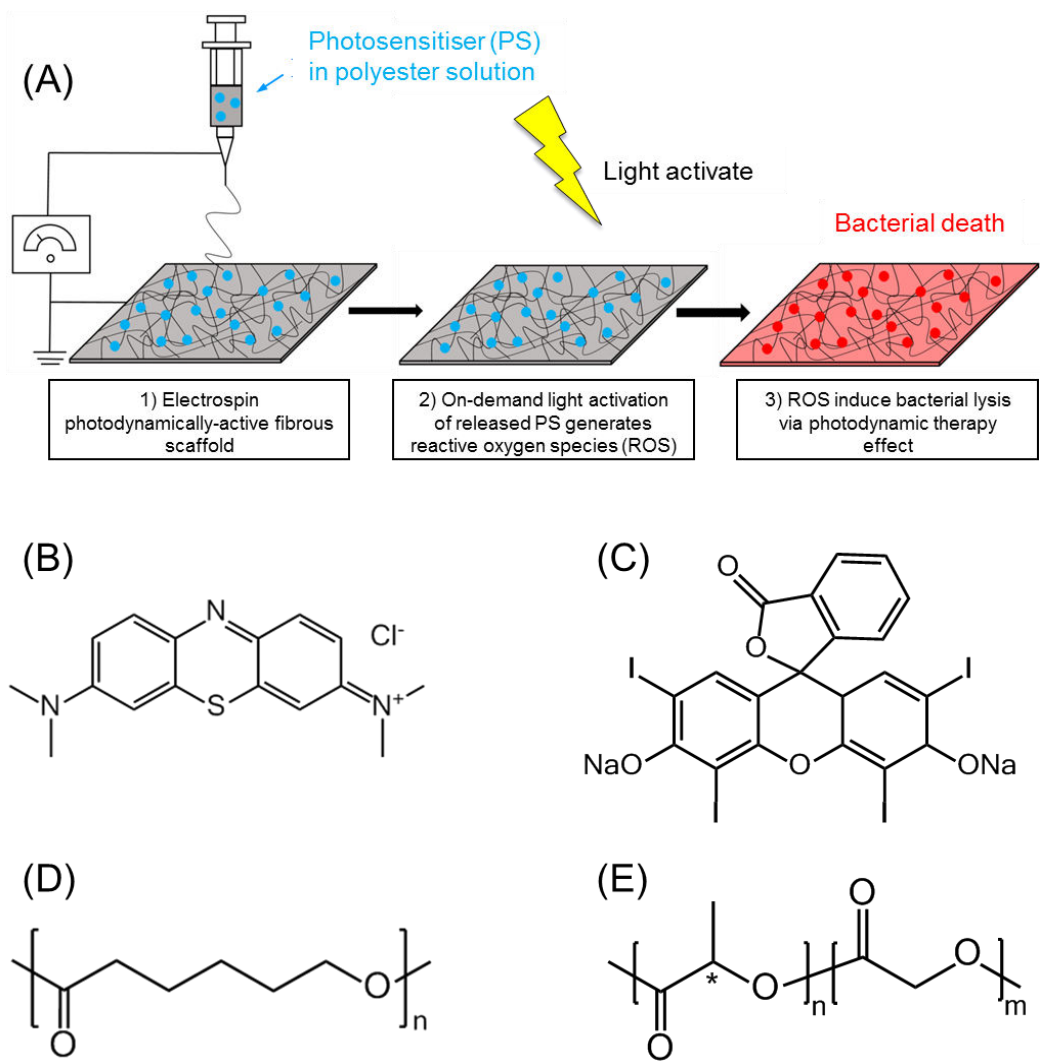

**Figure 1.** (A) Design and clinical applicability of photodynamically-active electrospun fibres for antibiotic-free infection control. (B-E): Chemical structure of selected PS and polymers used in this study. (B): Methylene Blue (MB); (C): Erythrosin B (ER); (D): Poly($\varepsilon$-caprolactone) (PCL). (E): Poly[(*rac*-lactide)-co-glycolide] (PLGA) with 75:25 monomer ratio.

## 3.1. Electrospinning Solution Characteristics

### 3.1.1. Characterisation of electrospinning solutions

The viscosity of the electrospinning solution is known to affect fibre formation and to alter the resulting diameter of the resulting electrospun fibres[30], ultimately impacting the PS release kinetics. Therefore, the viscosity of the electrospinning solutions both with and without PS for each polymer was determined. A commonly used volatile solvent to produce electrospinning solutions is hexafluoroisopropanol (HFIP), as it readily dissolves the polyesters and traces of the solvent can be removed from the finished product to safe levels with adequate drying[31]. Initial screening of the polymer solutions in HFIP was performed to gain comparable viscosities and spinnability between polymer groups. Other than the PS-free samples, a concentration of 2.2 mM of either MB or ER was employed in the electrospinning solutions, based on previous reports on MB and ER-induced aPDT[32,33] and aiming to achieve electrospun fibres with prolonged PS release and antimicrobial effect.

A typical shear thinning behaviour was observed in all electrospinning solutions regardless of the selected PS and polymer, whereby the solution viscosity was found to be inversely related to the shear rate, as expected for non-Newtonian liquids.

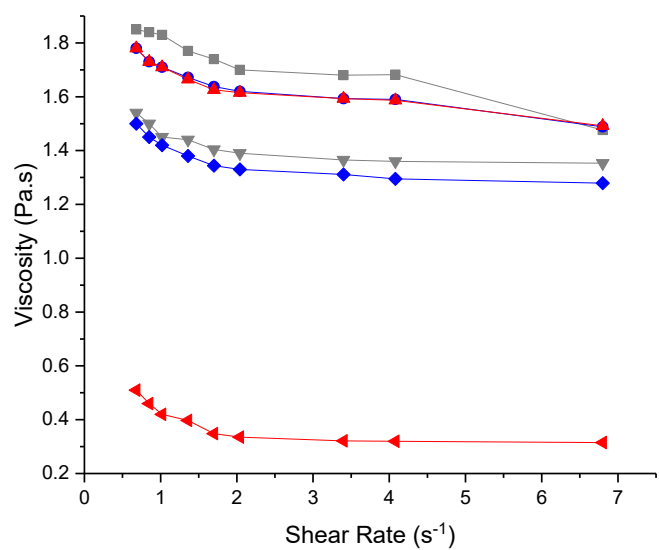

**Figure 2**. Viscosity of native and PS-loaded electrospinning polymer solutions prepared with either 6 wt.% PCL or 12 wt.% PLGA in HFIP. (■): PCL-ND; (●): PCL-MB; (▲): PCL-ER; (▼): PLGA-ND; (♦): PLGA-MB; (◄): PLGA-ER. Results reported as Mean±SD (n=3). Lines are guidelines to the eye.

By selecting polyesters with comparable molecular weight ($M_n$: 63,000-80,000 g·mol$^{-1}$), comparable viscosities were found between PCL-ND and PLGA-ND solutions at a shear rate of 6.8 s$^{-1}$ at concentrations of 6 wt.% PCL and 12 wt.% PLGA in HFIP ($\eta$ = 1.5 and 1.4 Pa·s respectively) (**Figure 2**).

Similar polymer concentrations have been reported for the formation of electrospun fibres with or without soluble factors[34,35]. Compared to respective PS-free polymer solutions, loading of PS did not induce detectable changes in the viscosity of the PCL-MB, PCL-ER or PLGA-MB polymer solutions (p= 0.10-0.12), whilst the viscosity of solution PLGA-ER proved to be significantly decreased (p=5.5x10$^{-16}$) (**Figure 2**). Previous studies reported that low concentrations of additives (< 12 mg·ml$^{-1}$) do not significantly change the viscosity of the electrospinning polyester solution[36], in agreement with the results obtained in this study. The significantly-decreased value of viscosity measured in PLGA-ER solutions with respect to solutions PLGA-ND and PLGA-MB may hint at secondary, e.g. hydrophobic, interactions between the PS and the fibre-forming polymer, as indirectly observed in ER-loaded PLGA nanoparticles[16]. Such secondary interactions between ER and PLGA are expected to compromise the polymer chain entanglements leading to a decrease in solution viscosity, as observed previously with different polymer and additive formulations[37,38].

The surface tension of solutions was also determined, since surface tension is expected to inversely relate to the electrospinnability of a given solution[39]. The surface tension appeared to be comparable between PCL- ($\sigma$ = 28±1–32±1 mN·m$^{-1}$) and PLGA-based ($\sigma$ = 32±1–33±2 mN·m$^{-1}$) electrospinning solutions, whilst the range of surface tension values was found to be in agreement with the one observed in previously-reported electrospinning polyester solutions[40]. There has been great interest into elucidating the relationship between surface tension and viscosity of electrospinning solutions and their effects on scaffold microarchitecture [41,42], since the fluid viscosity concerns the molecular interactions in the bulk of the solution, whereas the surface tension reflects the interactions of the solution at the air-liquid interface[41]. The above-mentioned surface tension results would therefore suggest that any change in the characteristics of the electrospun scaffolds are likely due to the PS-polymer-solvent secondary interaction in the bulk of the solution rather than at the air-liquid interface of the Taylor cone annd subsequent jets during electrospinning[43].

### 3.2. Electrospun Scaffold Characteristics

### 3.2.1. Scaffold Formation

Obtained polymer solutions successfully led to the formation of bead-free fibrous scaffolds (**Figure 3 A-F**), confirming that previously-measured solution viscosities and surface tensions were compatible with the electrospinning of selected polymers and PSs. To elucidate the scaffold loading efficiency and demonstrate the fibre encapsulation with either MB or ER, respective electrospun scaffolds were dissolved in HFIP to induce full release of incorporated PS. Photometric analysis of the resulting solution revealed a loading efficiency in the range of 97±30–110±16 wt.% (**Supporting Information Table S1**), therefore confirming that all the PS dissolved in the electrospinning solution was successfully encapsulated in the resulting fibres.

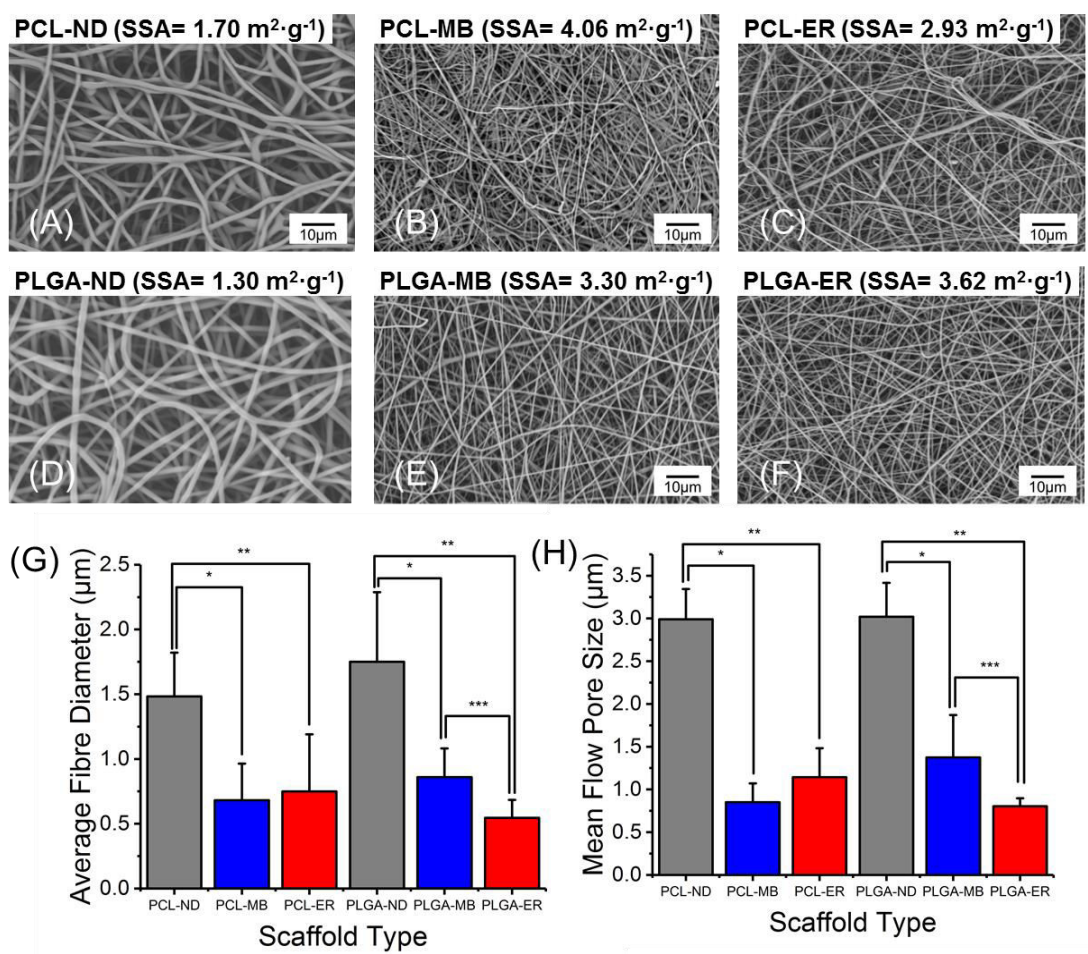

**Figure 3.** Microstructural analysis of PS-encapsulated scaffolds and electrospun controls. (A-F): SEM Images taken at 1000x magnification and specific surface area (SSA) measurements obtained via BET analysis. (G): Average fibre diameter determined from SEM images for each electrospun scaffold. (H): Mean flow pore size determined via porometry. '*', '**' and '***' denote significantly different means ($p$ <0.05, t-test).

Together with the photometric analysis, PS encapsulation proved to induce fibre colouration effects on the resultant scaffolds. Whilst PS-free and ER-encapsulated fibres appeared white- and red-like, respectively, regardless of the specific scaffold formulation, as expected, MB-incorporated scaffolds displayed either a purple- or blue-like colour depending on whether fibres were made of PCL or PLGA (**Supporting Information Figure S1A**). Although the colouration of PS-loaded materials is mainly determined by the specific PS and respective loading efficiency, as in the case of ER-encapsulated samples, the above-mentioned observations on MB-encapsulated samples suggest that secondary interactions between PS molecules and the polymer carrier may also play a role.

With regards to MB, it has been described in previous publications that loading of cellulosic derivative with MB species typically results in a blue colouration; however, when the PS concentration was increased, a purple colouration was observed in respective MB-encapsulated polymer, due to the aggregation of MB molecules via non-covalent pi-pi stacking interactions between aromatic rings of MB (**Supporting Information Figure S1B-C**) [44]. In this study, the higher viscosity measured in MB-loaded PCL solutions with respect to the corresponding PLGA variant (**Figure 2**) suggests a different state of MB molecules in the PCL solutions and resulting fibres. In the aggregated MB configuration, a lowered energy is required for the electrons to be excited, resulting in a red shift in the wavelength of visible light being absorbed and in a distinct fibre colouration effect.

### 3.2.2. Scaffold morphology

SEM, BET analysis and porometry were performed on the scaffolds resulting from electrospinning above-characterised polymer solutions, enabling quantification of fibre diameter as well as scaffold specific surface area and pore size, respectively. Despite employing the same molar concentration of PS, there was a significant reduction in fibre diameter upon encapsulation of either PS molecules in both scaffold systems (**Figure 3G**). For PCL scaffolds, encapsulation with either MB or ER resulted in 54 and 49% averaged reduction of fibre diameter, respectively, and similar values (51-69%) were also observed with PLGA-based samples. Such reduction of fibre diameter has been observed in other fibrous systems, deriving from electrospinning of PCL solutions containing peptides[45]. Introduction of ionically-charged PS, such as MB and ER, is likely to cause increased electrostatic repulsion between fibre-forming polymer jets in the electrospinning process[46]. For the PLGA-ER scaffolds, there was a further significant reduction in fibre diameter with respect to PS-free and MB-encapsulated PLGA scaffolds. This additional reduction in fibre diameter is in agreement with the significant decrease in viscosity observed in ER-loaded electrospinning solutions (**Figure 2**), since electrospinning solutions with reduced viscosity typically generate fibres with reduced diameter [47].

Porometry was next performed to determine the diameter of pores among the fibres within the fibrous structure (**Figure 3H and Supporting Information Figure S2**). The pore size between fibres is an important characteristic for a regenerative scaffold, as delivery of soluble factors, e.g. encapsulated PS, and cell infiltration have been shown to be altered by the pore size, with fibroblast cells being unable to bridge pores larger than 20 μm[48]. A pore size in the range of 0.7-3 μm was measured among the different scaffolds, whereby the scaffold formulation proved to induce variations in pore size comparable to those found with the fibre diameter, i.e. PS-encapsulated fibres were associated with scaffolds of decreased pore size. It is expected that an increased number of fibres with decreased diameter will be required to fill the same scaffold volume compared to fibres with an increased diameter. Previous porometry measurements therefore confirm the direct relationship between the fibre diameter and pore size in electrospun scaffolds[49]. The observed trends in fibre diameter is consistent with the variations in specific surface area of the scaffolds (**Figure 3 A-F**), since fibres with reduced diameter are expected to lead to scaffolds with increased specific surface area [50]. Overall, the averaged pore size was measured to be below 4 µm in all samples, suggesting that cell should be able to bridge these distances. Furthermore, this range of pore size is likely to promote the release of the PS molecule via a predominant diffusion mechanism through the scaffold, given the relatively low molecular weight of selected soluble factors ($M < 900$ g·mol$^{-1}$).

Interestingly, a narrow pore size distribution was found in PS-encapsulated scaffolds, in contrast to the broader range of pore size measured with the PS-free electrospun controls (**Supporting Information Figure S2**). Given the electrostatic charge of the PS molecules employed, this observation provides supporting evidence of the PS-induced electrostatic repulsion between polymer electrospinning jets. This ultimately results in PS-encapsulated scaffolds with more regular porous architectures with respect to the case of PS-free electrospun controls, as previously reported with other electrostatically-charged additives [46].

### 3.2.3. Tensile properties of electrospun scaffolds

The mechanical properties of the scaffolds were exemplarily measured on samples to investigate the potential effect of PS encapsulation on tensile properties (**Figure 4**). This investigation is important in order to explore the surgical handling capability and the scaffold applicability *in vivo*, e.g. for oral soft tissue applications, because the elasticity of the fibrous matrix has been shown to alter cell adhesion[51]. The mechanical properties of the scaffolds were measured on both PCL- and PLGA-based samples to investigate the potential effect of PS encapsulation on tensile properties.

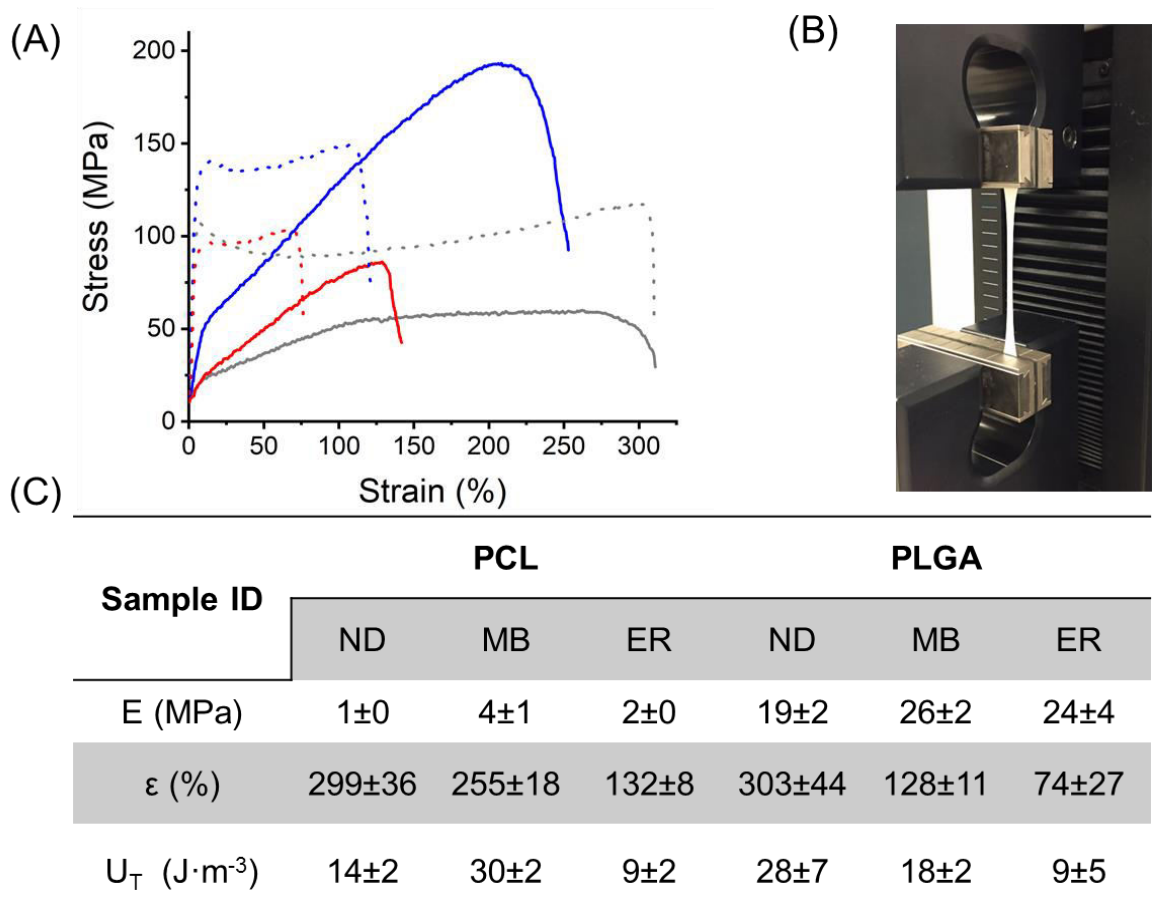

**Figure 4.** (A): Stress-strain curve of PS-encapsulated scaffolds and electrospun controls. (—): PCL-ND; (—): PCL-MB; (—): PCL-ER; (···): PLGA-ND; (···): PLGA-MB; (···): PLGA-ER. (B): Experimental setup employed during tensile testing. (C): Elastic modulus ($E$), strain at break ($\varepsilon$) and toughness ($U_T$) measured in PCL samples.

| Sample ID | PCL | | | PLGA | | |
|---|---|---|---|---|---|---|
| | ND | MB | ER | ND | MB | ER |
| $E$ (MPa) | 1±0 | 4±1 | 2±0 | 19±2 | 26±2 | 24±4 |
| $\varepsilon$ (%) | 299±36 | 255±18 | 132±8 | 303±44 | 128±11 | 74±27 |
| $U_T$ (J·m$^{-3}$) | 14±2 | 30±2 | 9±2 | 28±7 | 18±2 | 9±5 |

This investigation is also important in order to explore the surgical handling capability and the scaffold applicability *in vivo*, e.g. for oral soft tissue applications, because the elasticity of the fibrous matrix has been shown to alter cell adhesion[51]. When comparing the elastic modulus of PS-free control scaffolds, the PLGA7525-ND scaffold displayed significantly greater elastic modulus ($E$= 19±2 MPa, p=0.00005) with respect to scaffold PCL-ND ($E$= 1±0 MPa) (Figure 4C). This trend in mechanical properties between the two polyesters has been observed previously in the literature,[52] with comparable values of elastic modulus[53]. Following fibre encapsulation with the PS, both PCL-MB (*E*= 4±1 MPa) and PCL-ER (*E*= 2±0 MPa) scaffolds showed a significant modulus increase in comparison to PCL-ND controls (p= 0.00004 and 0.04 respectively), in line with other drug-loaded PCL scaffolds[54]. Similarly to the PCL-based samples, significant increase in elastic modulus was observed for PLGA7525-MB (*E*= 26±2 MPa, p=0.0004) versus PLGA7525, whilst the PLGA7525-ER scaffolds (*E*= 24±4 MPa) did not highlight any significant difference with respect to the control scaffolds (p=0.4).

The elastic modulus of the natural oral mucosa is thought to be approximately 3 MPa[55,56]. While the PCL scaffolds exhibited a comparable modulus to that of the native tissue, PLGA7525 scaffolds displayed increased tensile modulus with respect to the natural tissue. Consequently, the latter electrospun fibres may prove advantageous to enable easy surgical handling of the graft material during implantation minimising risks of material breakdown. The elasticity of the scaffold will also influence the interactions with contacting tissue in vivo[57].

Other than the elastic modulus, there was no significant difference in elongation at break ($\varepsilon$) values between the PCL-ND ($\varepsilon$= 299±36 %) and PLGA7525-ND ($\varepsilon$= 303±44 %) control scaffolds (p=0.9) (Figure 4C). Whilst, among the PCL-based samples, only the PCL-ER scaffolds showed a significant reduction in elongation at break ($\varepsilon$= 132±8 %, p=0.001) with respect to corresponding PS-free scaffold control, both PS-encapsulated PLGA scaffolds displayed a significant reduction in elongation at break (p= 0.0009 and 0.01 respectively) when compared to the PS-free control. These trends suggest that the polymer chains of the latter polyester could interact more strongly with selected PSs with respect to the polymer chains of PCL, likely due to different chemical structure, hydrophobicity and hydrogen bonding capability of respective polymer repeating units.

The impact of PS encapsulation on both electrospun systems is also reflected by the toughness values ($U_T$) obtained through integrating the area under each stress-strain curve (Figure 4C). The toughness value represents the energy required to fracture the material, so the higher the value, the tougher is the sample[58].. As with the elastic modulus, PLGA7525-ND samples displayed a significant increase in toughness ($U_T$= 28±7 J·m$^{-3}$) when compared to PCL-ND scaffolds ($U_T$= 14±2 J·m$^{-3}$, p=0.01). This increase in toughness and elasticity would make the polyester desirable for use as biomaterial scaffold.

For the PCL scaffolds, interestingly, there was a significant increase in toughness for the PCL-MB scaffolds ($U_T$= 30±3 J·m$^{-3}$, p=0.000009) but a significant decrease for PCL-ER scaffolds ($U_T$= 9±2 J·m$^{-3}$, p=0.01), with respect to the corresponding PS-free control. The increase in toughness and elasticity for the PCL-MB scaffolds is unexpected, as normally an increase in one of these properties reduces the other[59].. However, other reports of this phenomenon have been reported in previous literature with fibrous scaffolds[60].

There was no significant difference in the toughness of the PLGA7525-MB scaffolds ($U_T$= 18±2 J·m$^{-3}$, p=0.05) but the PLGA7525-ER scaffolds ($U_T$= 9±5 J·m$^{-3}$) displayed a significant decrease in toughness (p=0.01). Again, these changes in mechanical properties with the addition of soluble factors has been observed in previous studies[52], providing further indirect evidence of the development of secondary, e.g. hydrophobic, interactions between the PS molecule and the fibre-forming polymer.

## 3.3 Characterisation in aqueous environment

Following characterisation in the dry state, scaffolds were tested in contact with aqueous medium, in terms of wettability, degradability, PS release capability and aPDT effect.

### 3.3.1. Water contact Angle

The contact angle measurements allow quantification of the overall wettability of the scaffold. This is relevant since either PS diffusion or cell adhesion (to the surface of biomaterials) can be significantly affected by the surface wettability [61]. A contact angle of over 90° indicates a low interaction between the scaffold and the water (a hydrophobic response)[62]. Since fibre and pore size within the fibrous scaffold will affect the contact angle[63], water contact angle measurements were carried out on both the scaffolds and the pore-free films obtained via casting and drying of the same electrospinning solution (**Figure 5**).

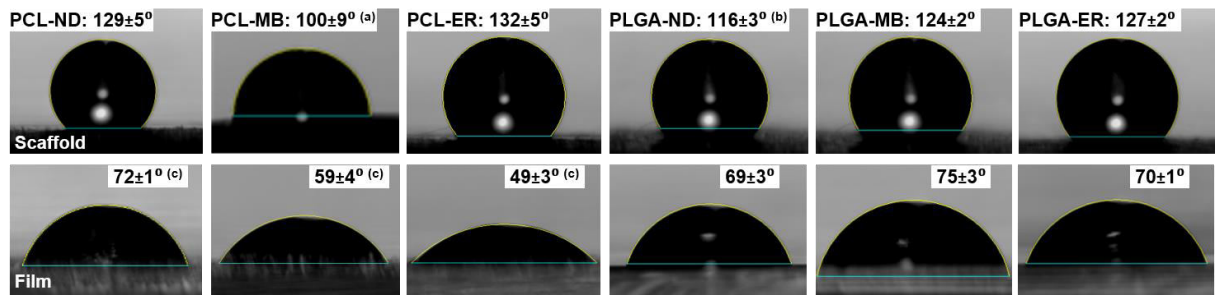

**Figure 5.** Water contact angle (WCA) images and measurements on dry PS-encapsulated and PS-free samples in the form of electrospun scaffold (top) and film (bottom). Data are presented as average ± standard deviation (n=3). (a)-(c): Statistical significance within either PCL or PLGA scaffold group (p<0.05, t-test).

Both PLGA and PCL scaffolds proved to display water contact angles higher than 90°, whereby a significant increase in contact angle was measured on PLGA scaffolds containing either MB or ER when compared to the PLGA-ND scaffold (P=0.03 and 0.008 respectively). An increase in contact angle has been measured on scaffolds with increased surface area[63], and this trend was confirmed in this study via BET analysis and porometry on PLGA scaffolds (**Figure 3A and 3C**). PCL-MB electrospun samples displayed a significantly-decreased water contact angle with respect to those of ER-encapsulated and PS-free electrospun samples, so the variation in contact angles across the different sample groups did not seem to agree with the effect of PS encapsulation on the surface of electrospun structures. In order to clarify this point, the effect of the fibre diameter and pore size was neglected, and pore-free films were analysed. Water contact angles well below 90° were measured in pore-free films obtained from respective electrospinning solutions, indicating that all films displayed increased compatibility with water in contrast to the case of the electrospun scaffold. There was a significant reduction in contact angle in both MB- and ER-loaded PCL films with respect to PS-free PCL controls (P=0.03 and P=0.005, respectively), in agreement with previous publications[45], whilst no significant difference was found for either the PLGA-MB or PLGA-ER films (P=0.06 and P=0.6 respectively) with respect to the PLGA-ND controls. Obtained contact angle values on PCL films therefore suggest that PS molecules directly interact with the water droplet, leading to increased wettability of the polymer surface with respect to the case of electrospun fibrous structures. Electrospun scaffolds are non-homogeneous materials made of solid fibres and pores, such that the superficial discontinuities (i.e. air) are responsible for the different wetting behaviour of scaffolds with respect to the case of pore-free samples. On the other hand, film formation is a different process to that of electrospinning, whereby the polymer solution (with or without PS) is cast and air-dried.

### 3.3.2. Photosensitiser release

Following confirmation of PS encapsulation in the scaffolds, electrospun samples were incubated in aqueous medium and macroscopic behaviour and PS release monitored. None of the PS-encapsulated PCL samples showed a significant change in dimensions during the selected incubation time, whilst a drastic macroscopic shrinkage was observed with respective PLGA variants upon contact with water (**Supporting Information Figure S3 and S4**). PS-encapsulated PLGA samples reduced in macroscopic surface area by approximately 50% when compared to the PLGA-ND scaffolds. Such variation in macroscopic dimensions is likely explained by the fact that PLGA fibres display an amorphous polymer morphology; consequently, water molecules can access relatively freely throughout the polymer chains, acting as plasticiser and inducing increased chain mobility [64]. In contrast, the crystalline regions in PCL fibres present limited accessibility to water molecules, therefore acting as physical crosslinks and preventing volumetric change in hydrated scaffold dimensions. The water-induced plasticising effect of PLGA fibres is dominant in PS-encapsulated samples, given that respective fibres proved to display a significantly-decreased diameter with respect to the case of electrospun control fibres (51% and 69% reduction respectively). In line with the PS-induced decrease of fibre diameter and increased water uptake, merging of fibres and collapse of the porous scaffold architecture are increasingly likely.

Other than the minimal macroscopic variations in hydrated conditions, electrospun PCL samples generally described a faster PS release profile with respect to respective PLGA variants, whilst MB proved to be more readily released compared to ER, regardless of the polymer carrier employed (**Figure 6**).

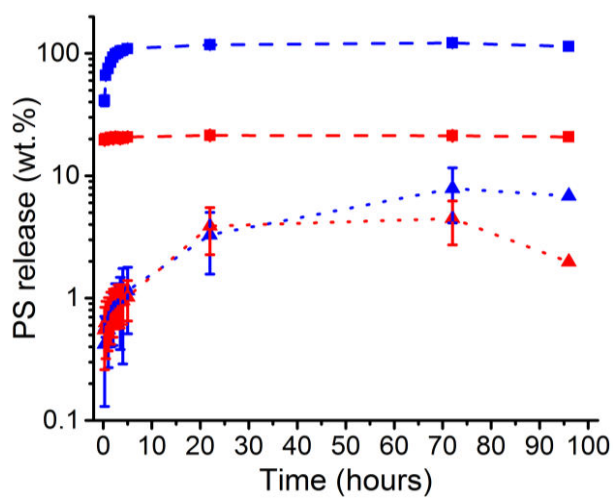

**Figure 6.** Typical PS release profiles measured via UV-Vis spectroscopy of the supernatant collected during incubation ($H_2O$, 37 $^{0}C$) of MB- and ER-encapsulated scaffolds at selected time points. (-■-): PCL-MB; (-■-): PCL-ER; (··▲··): PLGA-MB; (··▲··): PLGA-ER. Lines are guidelines to the eye.

Scaffolds PCL-MB displayed complete release following 96-hour incubation, whereby the cumulative mass of MB measured in the supernatant ($m$: 170±9 mg) was found to compare well with the mass of PS encapsulated in the electrospun fibres ($m$: 150±2 mg). In comparison, only a limited amount of PS was released from both PCL-ER and the PLGA scaffolds, suggesting that PS is being held within the scaffold either due to secondary interactions between the PS molecule and the polymer as well as due to the macroscopic shrinkage of the PLGA scaffolds.

The different PS release profiles recorded from PCL and PLGA electrospun scaffolds were found to be somewhat surprising, given that the averaged pore size was comparable between the two scaffold architectures (**Figure 3H**) and that the amorphous morphology of PLGA should allow for increased diffusion of the PS out of the electrospun fibres with respect to semi-crystalline PCL. The most likely explanation for the increased release capability of PCL with respect to PLGA samples is that the encapsulation of PS in the PCL fibres leads to increased surface hydrophilicity, as demonstrated by contact angle data (**Figure 5**), so that diffusion of PS molecules is promoted in the PCL samples. The higher and faster release of MB with respect to ER can on the other hand be explained considering the different solubility in water and molecular weight of the two PS. MB (*solubility*= 35.5 mg·ml$^{-1}$; *M*= 319.85 g·mol$^{-1}$) is more soluble in aqueous envinroment with respect to ER (*solubility*= 0.7 mg·ml$^{-1}$; *M*= 879.86 g·mol$^{-1}$), so that an increased diffusion of MB out of the scaffold is expected. Overall, the burst release observed with all samples is commonly seen with fibrous scaffolds used in drug delivery applications [65]. A steady PS release would be preferred to allow repeated activation of the PS to treat infections which may arise. Altering the monomer ratios in PLGA polymers could open up relevant avenues to induce polymer crystallisation enabling dimensional stability, on the one hand, and weak PS-polymer interactions, on the other hand. It is expected that a steady release could also be achieved by building fibrous configurations integrating PS-loaded fibres with outer PS-free scaffolds acting as barrier minimising burst release.

### 3.3.3. Water Uptake and hydrolytic degradability

As the hydrolysis of polyester is a second order reaction, the reaction rate of hydrolysis will be dependent upon the water uptake and swelling of the polymer with water [66]. Other than hydrolysis, the water uptake into the scaffold will also affect the scaffold release capability as well as cellular tolerability and regenerative potential *in vivo*. Following 24-hour incubation in aqueous medium, a significantly greater water uptake was measured in PCL-ND compared to PLGA-ND scaffolds (**Supporting Information Figure S5**), and the same trends were observed in respective PS-encapsulated samples. Although detectable release of PS was measured within the selected water uptake time window (**Figure 6**), the mass percentage loss of PS in the electrospun fibre was minimal with respect to that of the control polymer fibres, suggesting that the measured water uptake was mostly ascribed to the effect of the scaffold chemical composition and architecture rather than the diffusion of PS out of the material.

Whilst the averaged fibre diameter and mean flow pore size measured in PCL and PLGA sample groups were statistically equivalent (**Figure 3G-H**), the water uptake results are in agreement with previous contact angle (**Figure 5**) and PS release (**Figure 6**) measurements, indicating a higher compatibility with water in PCL with respect to the PLGA samples. The water uptake measured on PS-encapsulated PCL scaffolds was greater than the PLGA samples and electrospun controls. These trends in water uptake are found to correlate with the decrease in fibre diameter and pore size and the increase in hydrophilicity observed in the water contact angle experiments on the films recorded in both samples PCL-MB and PCL-ER (**Figure 3G-H**) with respect to sample PCL-ND.

Degradability of regenerative devices needs to be tailored for the intended tissue repair/clinical application, therefore the hydrolytic degradability of each fibrous system in this study was investigated [34]. Previous research performed on porcine palatal wounds found that full clinical closure of the small wounds had occurred by 14 days, and complete healing of the wound had occured after 7 weeks[67]. This timescale will change depending on the size of the wound, and may also differ in humans. It was therefore determined that the scaffolds would need to maintain integrity for 6-8 weeks in the oral cavity to allow for support of neotissue formation to occur. Changes in scaffold microarchitecture (**Figure 7** and **Supporting Information**

**Figure S6**), macroscopic volume (**Supporting Information Figure S3 and S4)** and sample mass (**Supporting Information Figure S7**) were monitored following sample incubation in PBS for up to 8 weeks.

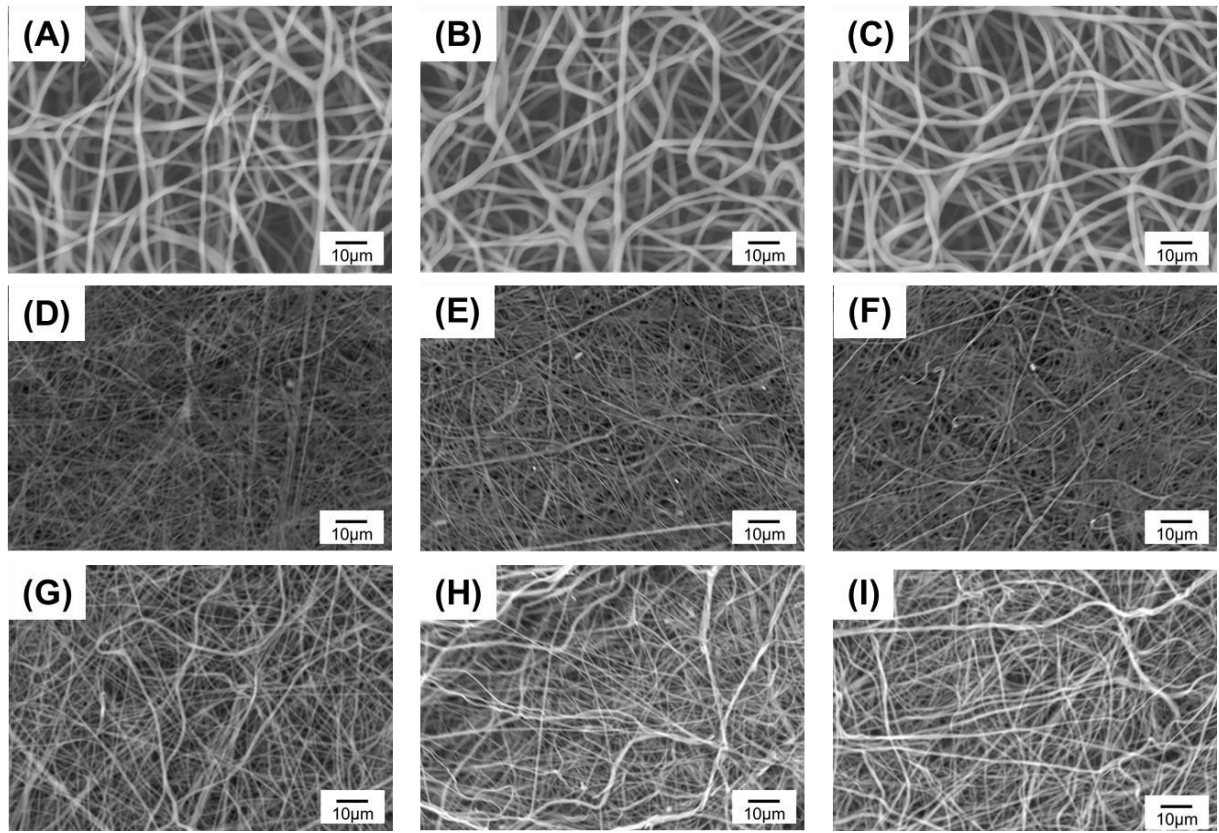

**Figure 7.** Scanning Electron Microscopy (SEM) of electrospun PCL scaffolds following 8-week incubation in PBS at 37 $^0$C. (A-C): samples PCL-ND following 1 (A), 4 (B) and 8 (C) weeks. (D-F): samples PCL-MB following 1 (D), 4 (E) and 8 (F) weeks. (G-I): samples PCL-ER following 1 (G), 4 (H) and 8 (I) weeks.

Minimal structural changes were observed for the retrieved PCL scaffolds at all selected time points (**Figure 7**), whilst both PLGA-ND and PLGA-MB scaffolds revealed a decrease in pore size between fibres after 1 week incubation in PBS (**Supporting Information Figure S6**). Following 8 weeks, PLGA-MB scaffolds were disintegrated and a collapsed fibrous architecture was observed, in line with the previously-observed macroscopic volume reduction (**Supporting Information Figure S4**). In line with the greater instability at both macroscopic and microscopic scales, electrospun PLGA scaffolds proved to display a higher mass loss (14±4 wt.%) than PCL scaffolds (4±2 wt.%) following 8 weeks (**Supporting Information Figure S7**). Given the increased hydrolytic degradability of PLGA with respect to PCL, an increased PS release should be observed in the former with respect to the latter scaffolds, following polymer hydrolysis and breakdown of the scaffold.

Lowery *et al.* has previously published research which concludes that the pore size significantly alters cell adhesion within electrospun scaffold[48]. As the scaffold becomes non-porous in a moist environment, cells would not be able to inflitrate so the scaffold would not function well as a regenerative device. For this reason, the PLGA formulations were withdrawn from further studies.

### 3.3.4 Antibacterial activity

The PCL scaffolds were selected over PLGA scaffolds for evaluation of their aPDT effect and bactericidal capability (**Figure 8**), due to their dimensional stability in hydrated environment. The aPDT effect has already been shown to be effective against Gram-positive bacteria [68]; consequently, *E.coli* was used in our studies to test the effect of aPDT on a model Gram-negative bacteria. The scaffolds were incubated in medium for 2 hours before bacteria was added directly onto the scaffold with light exposure to reflect clinical applications because most likely the scaffolds would be implanted for a period of time prior to the detection of an infection. To verify that the selected PS molecules were not toxic prior to light activation, an identical set of foil-covered scaffold samples were analysed in each experiment to calculate the 'dark toxicity' of the scaffolds.

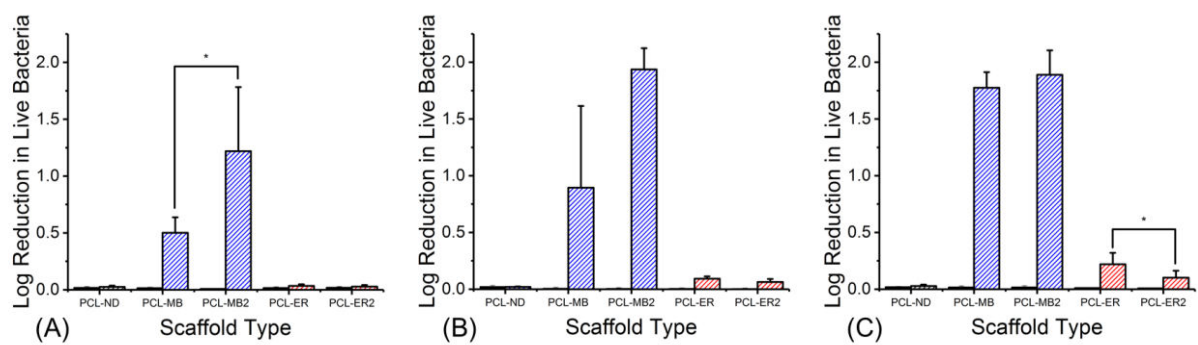

**Figure 8.** Average log reduction of live bacteria cultured on PS-loaded PCL scaffolds following 30-min (A), 60-min (B) and 120-min (C) light exposure. Scaffolds PCL-MB2 and PCL-ER2 were electrospun from polymer solutions with doubled PS concentration ([PS]= 4.4 mM). Black bars refer to data obtained following incubation in dark. Hashed bars represent light measurements. Results are reported as Mean±SD. '*' denotes significantly different means ($p < 0.05$, t-test).

In all experiments, the dark toxicity of the PS-encapsulated scaffolds was below 0.05 log reduction (**Figure 8**), and comparable values were measured with PS-free electrospun scaffold controls. A key observation observed across all PS-containing scaffold formulations, was that the longer the light exposure, the more bacteria were killed, e.g. for PCL-MB scaffolds following 30, 60 or 120 minute light exposure, the log reductions were 0.5, 0.9 and 1.8 respectively. Our finding is in agreement with previous reports that the greater the dose of light, the more toxic reactive species are generated and the more active the PS is[14]. Other than that, MB-encapsulated scaffolds with the same PS concentration were found to kill significantly more *E. coli* than ER-encapsulated scaffolds. This observation is in agreement with the increased release of MB from the scaffold with respect to ER (**Figure 6**), so that an increase in cellular MB uptake is expected. The increased aPDT effect of MB-encapsulated fibres may also due to the fact that MB is cationic, which would result in greater interaction with the Gram-negative *E.coli* cell membranes[9].

Scaffolds electrospun with twice the MB concentration (4.4 mM) did not increase the killing of microbes after 60 minutes light exposure. This confirms that the originally-selected MB dose (2.2 mM) is suitable to trigger significant aPDT effect. An unexpected observation was that after 120 minutes, there was a small but significant difference between ER double strength and ER single strength, with the double strength ER killing significantly less in certain comparable experiments (p=0.04). One possible explanation is that the ER molecule is released more quickly from the double strength scaffolds during the pre-incubation stage in medium, so that the effectiveness of the light-induced PS activation reaction is compromised. Another possible explanation would be that the ER was not fully soluble at this concentration and therefore all of the ER molecules were not incorporated into the scaffold.

## 4. Conclusions

This study has successfully demonstrated the potential of a prototype regenerative medical device integrated with aPDT capability. Either MB or ER encapsulation into and release from fibrous electrospun scaffolds have been achieved with two FDA-approved fibre-forming polyesters, i.e. PCL and PLGA. Despite only one electrospinning solution formulation (PLGA-ER) displaying a significantly decreased viscosity, all PS-encapsulated scaffolds exhibited significantly reduced average fibre diameter and pore size. In physiological environment, PLGA scaffolds displayed significant shrinkage upon contact with water and collapse of the fibrous structure following 8-week incubation, likely explained by the amorphous polymer morphology of PLGA with respect to the semi-crystalline state of PCL fibres.

This shrinkage resulted in the PLGA scaffolds being deemed not suitable for use in the moist environment of the mouth, due to reduced porosity which would prevent cell infiltration and therefore neotissue formation. Together with the hydrolytic degradation study, PCL scaffolds were deemed to be a dimensionally-stable polymer carrier for both PS molecules, and MB-loaded PCL scaffolds showed to be most effective at killing bacteria with selected PS concentrations. Therefore, the PCL-MB scaffold system represents the most suited prototype for the intended application in the presented study. An important consideration for the use of this scaffold in dental surgery is the source of light, since this should be selective and powerful enough to enable prompt antimicrobial effect. Although an LED light was employed in this study so that at least 30-min irradiation was required to induce aPDT effect, the use of a dental laser is expected to be compatible with clinically-relevant irradiation times (ideally no longer than 10 minutes) in order to activate the PS, since the light intensity is increased. Future work from this study should address the antibacterial capability of these scaffolds against anaerobic oral bacteria, particularly in biofilm state, in light of their resistance to conventional scaffold-free aPDT[69].

**Supporting Information:**

Table S1 - Loading efficiency (LE) and percent release measured in PCL and PLGA scaffolds electrospun in the presence of either MB or ER

Figure S1 - (A) Macroscopic images of PS-free and PS-encapsulated scaffolds. (B) Aggregation of MB molecules results in a purple colour of PS-encapsulated fibres. (C) Encapsulation of MB in the monomeric state results in a blue colour of respective fibres.

Figure S2 - Typical pore size flow distribution measured via porometry in electrospun scaffolds.

Figure S3 - Macroscopic images of electrospun PCL scaffolds following electrospinning (A-C) and 8-week hydrolytic incubation (D-F) in 37 $^0$C distilled water.

Figure S4 Macroscopic images of electrospun PLGA scaffolds following electrospinning (A-C) and 8-week hydrolytic incubation (D-F) in 37 $^0$C distilled water.

Figure S5 - Water uptake measured gravimetrically following incubation ($H_2O$, 37 $^0$C) of either PS-loaded or electrospun control samples.

Figure S6 - Scanning Electron Microscopy (SEM) of electrospun PLGA scaffolds following 8-week hydrolytic incubation (PBS, 37 $^0$C).

Figure S7 - Mass loss measured on samples PCL-ND (black) and PLGA-ND (grey) following hydrolytic degradation ($H_2O$, 37 $^0$C).

**Acknowledgements**

This research was funded for by the Engineering and Physical Research Council and iCASE PhD with industry sponsor Neotherix. Ltd.

**Declaration of Interest**

Michael J. Raxworthy is the CEO of Neotherix Ltd.

# Supporting Information

**Photodynamically Active Electrospun Fibres for Antibiotic-Free Infection Control**

Amy Contreras,[1] Mike Raxworthy,[2] Simon Wood,[3] Jessica D. Schiffman,[4] Giuseppe Tronci[3,5] *

[1] Institute of Medical and Biological Engineering, University of Leeds, Leeds, LS2 9JT, UK (mnasm@leeds.ac.uk)

[2] Neotherix Ltd., The Hiscox Building, Peasholme Green, York, YO1 7PR, UK (mike.raxworthy@neotherix.com)

[3] School of Dentistry, University of Leeds, Leeds, LS2 9JT, UK (s.r.wood@leeds.ac.uk)

[4] Department of Chemical Engineering, University of Massachusetts Amherst, 240 Thatcher Rd, Amherst MA 01003-9364, USA (schiffman@ecs.umass.edu)

[5] School of Design and School of Dentistry, University of Leeds, Leeds, LS2 9JT, UK (g.Tronci@leeds.ac.uk)

* Email correspondence: g.tronci@leeds.ac.uk (G.T.)

**Table S1.** Loading efficiency (*LE*) and percent release measured in PCL and PLGA scaffolds electrospun in the presence of either MB or ER.

| Sample ID | PCL-MB | PCL-ER | PLGA-MB | PLGA-ER |
|---|---|---|---|---|
| *LE* /wt.% | 103±16 | 103±31 | 110±16 | 97±30 |
| *% release* /wt.% [1] | 114±4 | 28±2 | 7±4 | 2±1 |

[1] Percent release following 8-week sample incubation (PBS, 37 $^0$C).

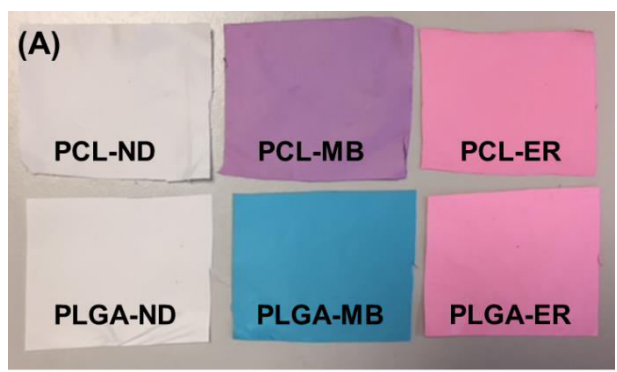

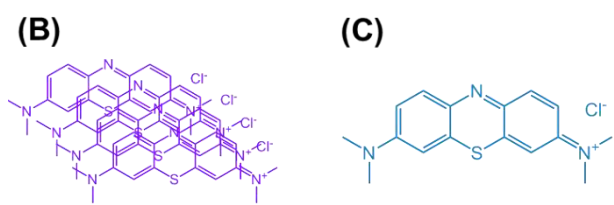

**Figure S1.** (A) Macroscopic images of PS-free and PS-encapsulated scaffolds. (B) Aggregation of MB molecules results in a purple colour of PS-encapsulated fibres. (C) Encapsulation of MB in the monomeric state results in a blue colour of respective fibres.

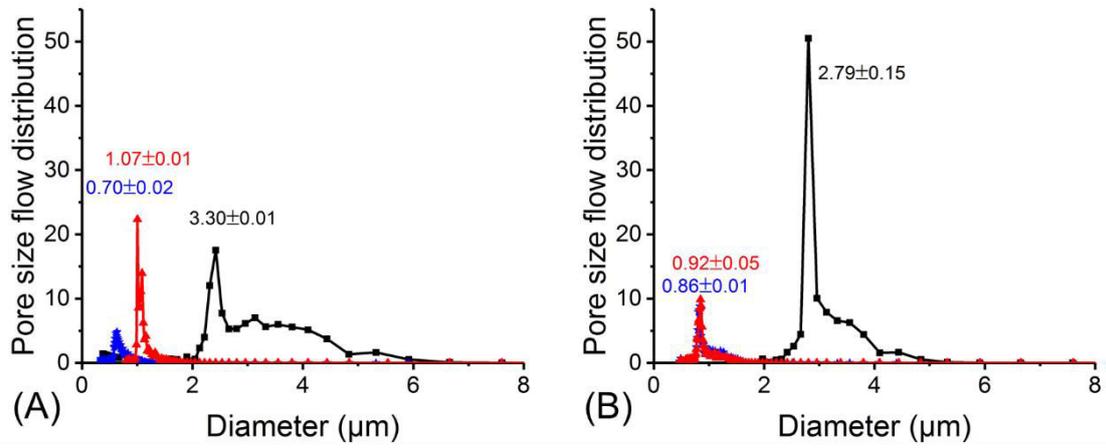

**Figure S2.** Typical pore size flow distribution measured via porometry in electrospun scaffolds of PCL (A) and PLGA (B). (■): PS-free (ND); (★): MB-encapsulated; (▲): ER-encapsulated.

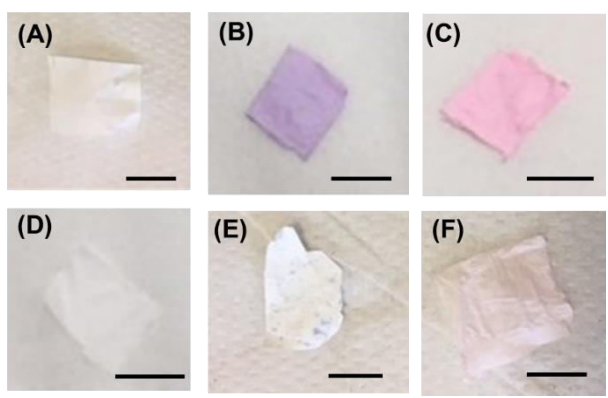

**Figure S3.** Macroscopic images of electrospun PCL scaffolds following electrospinning (A-C) and 8-week hydrolytic incubation (D-F) in 37 $^0$C distilled water. (A, D): PCL-ND; (B, E): PCL-MB; (C, F): PCL-ER. Scale bar: ~ 1 cm.

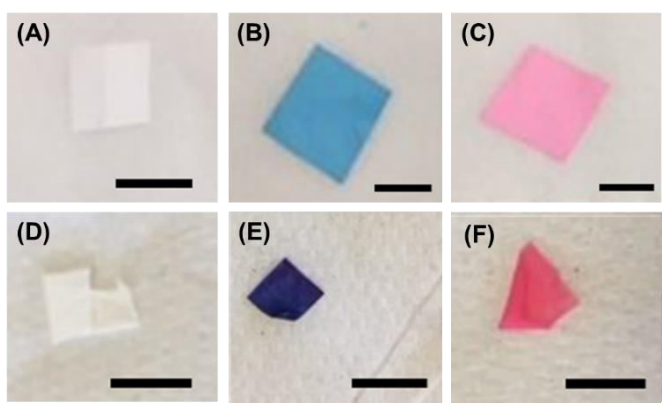

**Figure S4.** Macroscopic images of electrospun PLGA scaffolds following electrospinning (A-C) and 8-week hydrolytic incubation (D-F) in 37 $^0$C distilled water. (A, D): PLGA-ND; (B, E): PLGA-MB; (C, F): PLGA-ER. Scale bar: ~ 1 cm.

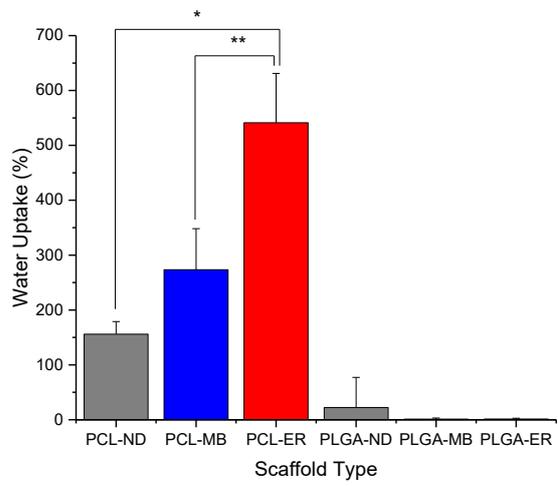

**Figure S5.** Water uptake measured gravimetrically following incubation ($H_2O$, 37 $^{\circ}$C) of either PS-loaded or electrospun control samples. '*' and '**' denote significantly different means ($p$ <0.05, t-test).

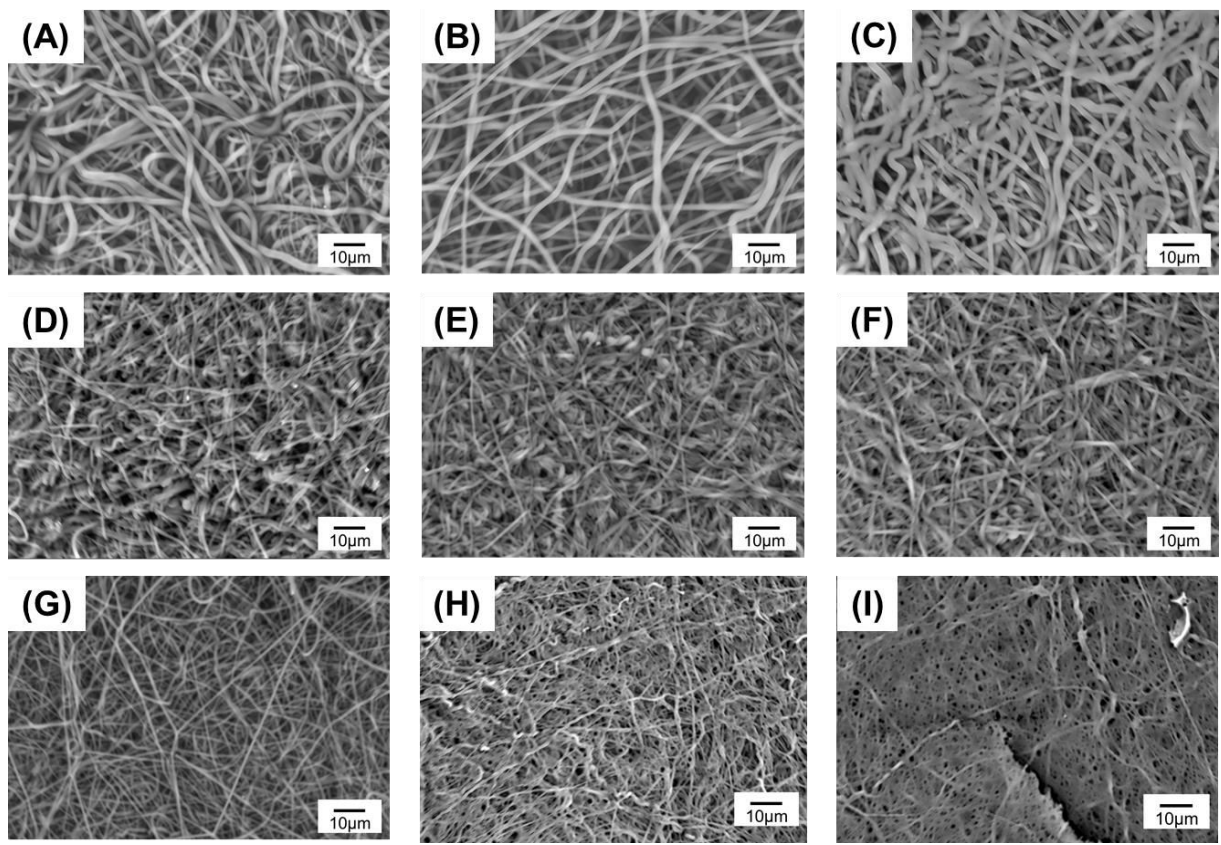

**Figure S6.** Scanning Electron Microscopy (SEM) of electrospun PLGA scaffolds following 8-week hydrolytic incubation (PBS, 37 $^{o}$C). (A-C): samples PLGA-ND following 1 (A), 4 (B) and 8 (C) weeks. (D-F): samples PLGA-MB following 1 (D), 4 (E) and 8 (F) weeks. (G-I): samples PLGA-ER following 1 (G), 4 (H) and 8 (I) weeks.

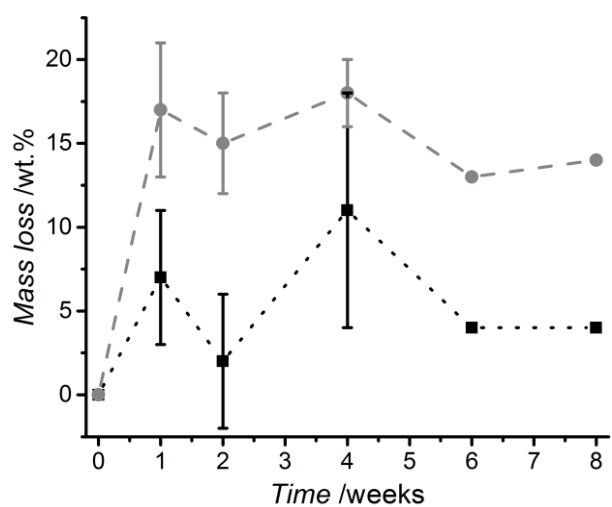

**Figure S7.** Mass loss measured on samples PCL-ND (black) and PLGA-ND (grey) following hydrolytic degradation (H$_2$O, 37 $^0$C). Lines are guidelines to the eye.